\documentclass[preprint,aps,prd,showpacs,nofootinbib,superscriptaddress,tightenlines]{revtex4}

\usepackage{amsfonts}
\usepackage{multirow}
\usepackage{mathrsfs}
\usepackage{graphicx}
\usepackage{amsmath}
\usepackage{amssymb}
\usepackage{bm}
\usepackage{bbm}

\def\ups{\Upsilon}
\def\jpsi{J/\psi}
\def\psis{J/\psi}
\def\etac{\eta_c}
\def\ae{\alpha}
\def\as{\alpha_s}

\def\eps{\epsilon}
\def\PA#1{\left( #1 \right)}
\def\PB#1{\left[ #1 \right]}

\def\v2#1{\langle v^2\rangle_{#1}}
\def\nn{\nonumber}

\newcommand{\beq}{\begin{equation}}
\newcommand{\eeq}{\end{equation}}
\newcommand{\bqa}{\begin{eqnarray}}
\newcommand{\eqa}{\end{eqnarray}}

\begin{document}
\title{\mbox{}\\[10pt]
Relativistic corrections to $\Upsilon$ exclusive decay into double $S$-wave charmonia}

\author{Wen-Long Sang} 
\affiliation{School of Physical Science and Technology, Southwest
University, Chongqing 400700, China\vspace{0.2cm}}
\affiliation{State Key Laboratory of Theoretical Physics, Institute of Theoretical Physics, Chinese Academy of Sciences,
Beijing 100190, China\vspace{0.2cm}}

\author{Feng Feng}\footnote{F.Feng@outlook.com}
\affiliation{China University of Mining and Technology, Beijing 100083, China\vspace{0.2cm}}
\affiliation{State Key Laboratory of Theoretical Physics, Institute of Theoretical Physics, Chinese Academy of Sciences,
Beijing 100190, China\vspace{0.2cm}}

\author{Yu-Qi Chen}
\affiliation{State Key Laboratory of Theoretical Physics, Institute of Theoretical Physics, Chinese Academy of Sciences,
Beijing 100190, China\vspace{0.2cm}}
\date{\today}

\begin{abstract}
Within the framework of nonrelativisitic QCD factorization formalism, we present the next-to-leading-order relativistic corrections to $\Upsilon$ exclusive decay into $\eta_c$ plus $J/\psi$.
The double charmonia can be produced through several immediate channels, i.e.,
$\ups\to g^*g^*g^*\to \etac+\psis$, $\ups\to g^*g^*\gamma^*\to \etac+\psis$, and $\ups\to \gamma^*\to \etac+\psis$. The amplitudes of these three channels are obtained accurate up to ${\cal O}(\as^3 v^2)$,
${\cal O}(\ae\as^2 v^2)$, and ${\cal O}(\ae\as^2 v^2)$, respectively, where $v$ indicates the typical heavy quark velocity in bottomonium and/or charmonium rest frame. The decay rates are also presented. We find that the next-to-leading-order relativistic corrections to the short-distance coefficients as well as the decay rates are both significant and negative, especially for the corrections from bottomonium. More seriously, the decay rates are even brought into negative by including the relativistic corrections, which indicates the poor convergence for the velocity expansion in this kind of process. Detailed analysis is given in the paper.
\end{abstract}


\maketitle

\section{Introduction}

Nonrelativistic QCD (NRQCD) \cite{Bodwin:1994jh} is a powerful and successful effective field theory in describing the quarkonium production and decay.
The processes of bottomonium exclusive decay into double charmonia are interesting channels to investigate quarkonium, which
have been extensively studied both in theory and experiment~\cite{Shen:2012ei,Yang:2014yyy}. The $\eta_b$ decay into double $J/\psi$ was studied
in Refs.~\cite{Jia:2006rx,Santorelli:2007vm,Santorelli:2007xg,Santorelli:2008mn,Braguta:2009xu},
and the radiative and relativistic corrections to this process were
investigated subsequently in \cite{Gong:2008ue,Sun:2010qx,Sang:2011fw}.
The $P$-wave bottomonium exclusive decay into double charmonia was also studied in Refs.~\cite{Braguta:2005gw,Braguta:2009df}. The relativistic corrections to these processes were
obtained in Refs.~\cite{Sang:2011fw,Zhang:2011ng}, and the radiative corrections to
$\chi_{bJ}\to J/\psi J /\psi$ were calculated recently \cite{Chen:2014lqa}.
For the case of $\ups$, the $\ups\to \etac J/\psi$ and $\ups\to J/\psi\chi_{cJ}$ were separately studied in Ref.~\cite{Jia:2007hy} and Ref.~\cite{Xu:2012uh}.

Comparing with other bottomonium exclusive decay processes mentioned above, $\ups\to \etac J/\psi$ has some distinctive features. First, there is no tree-level pure QCD Feynman diagram. Second, this process
is further suppressed by a helicity selection rule due to $m_b\gg m_c$. As a consequence,
one may expect that the decay rate is relatively small~\cite{Jia:2007hy}. Nevertheless,
it is still meaningful to investigate this process based on the following reasons. First, there is little study of the relativistic corrections to the quarkonium decay and production processes with the vanishing tree-level Feynman diagram, so it is theoretically interesting to study the relativistic corrections and the convergence of the velocity expansion for these processes. Second,
it has been known for a long time that the relativistic corrections to $J/\psi\to {\rm LH}$ and $J/\psi\to 3\gamma$ are
both significant and negative~\cite{Keung:1982jb}, which even surpasses the contributions at leading order.
The study in Ref.~\cite{Feng:2012by} indicated that the ${\cal O}(\alpha_s v^2)$ corrections to $J/\psi\to 3\gamma$
 are also huge; meanwhile, the study in Refs.~\cite{Chen:2011ph,Chen:2012zzg} indicates that the relativistic
corrections to $\ups\to g^*gg\to c\bar{c}gg$ are extraordinarily large, despite the small
${\cal O}(v^2)$ NRQCD matrix element for $\ups$. The common feature for these processes is that $\ups$ first annihilates into three gluons (photons).
For the process $\ups\to \etac J/\psi$, the main contributions come from the immediate channel
$\ups\to g^*g^*g^*\to \etac J/\psi$~\cite{Jia:2007hy}. As a consequence, it is compelling to know
whether large relativistic corrections are also appearing in $\ups\to g^*g^*g^*$. Third,
there has been a longstanding ``$\rho$-$\pi$" puzzle in quarkonium decay~\cite{Agashe:2014kda,Brambilla:2010cs}.
The structure of Feynman diagrams for $\psi\to \rho\pi$ is rather similar to $\ups\to \etac J/\psi$,
so one may wonder whether the puzzle originates from the relativistic corrections for $\psis$.
Based on these reasons, we will investigate the relativistic corrections to $\ups\to \etac J/\psi$ from both the
charmonia and the bottomonium.

Up to the next-to-leading order in velocity expansion, there are three channels which contribute,
$\ups\to g^*g^*g^*\to \etac\psis$, $\ups\to g^*g^*\gamma^*\to \etac\psis$, and $\ups\to \gamma^*\to \etac\psis$. The leading-order (LO) amplitudes for the three channels in powers of $\as$ and
$\ae$ scale as $\as^3$, $\alpha\as^2$, and $\alpha\as$, respectively. Though the second channel is
suppressed by a factor $\alpha/\as$ relative to the
first channel, there is a fragmentation enhancement ($\gamma^*\to J/\psi$) in
$\ups\to g^*g^*\gamma^*\to \etac\psis$; as a result, we also consider this channel with $J/\psi$ produced by a photon in our calculation. Moreover, since
$\ups\to \gamma^*\to \etac\psis$ is of a similar Feynman diagram structure as $e^+e^-\to \etac J/\psi$, the
radiative corrections to which were found to be large~\cite{Zhang:2005cha,Gong:2007db}, the calculation in
this channel will be accurate up to ${\cal O}(\alpha\as^2 v^2)$.~\footnote{We do not consider the pure QED contributions in $\ups\to \gamma^*\to \etac\psis$, due to the fact
the pure QED contributions are less than a third of other contributions~\cite{Braaten:2002fi}, and what is more, $\ups\to \gamma^*\to \etac\psis$
contributes merely a small fraction in $\ups\to\etac\psis$ decay~\cite{Jia:2007hy}.}

The rest of the paper is organized as follows. In Sec.~\ref{sec-NRQCDformula}, we describe the NRQCD formula for
$\ups$ exclusive decay into double charmonia and present the kinematic descriptions and required techniques.
The short-distance coefficients (SDCs) are obtained in detail in Sec.~\ref{sec-amplitudedetail}. Section \ref{sec-prediction} is devoted to
numerical predictions. Discussions and summary are also presented in this section.

\section{NRQCD formula and kinematic descriptions\label{sec-NRQCDformula}}

According to Lorentz invariance, we are able to factorize $\Upsilon$ exclusive decay into double $S$-wave charmonia as
\beq{\label{eq-formfactor}}
{\cal A}\equiv\langle\etac\psis(\eps_2)|\ups(\eps_1)\rangle=i\epsilon^{\mu\nu\alpha\beta}
\epsilon_{1\mu}\epsilon_{2\nu}^*P_{1\alpha}P_{2\beta}A,
\eeq
where $\eps$ tensor indicates the Levi-Civita symbol, $\eps_1$ and $\eps_2$ represent the polarization vectors of the initial state $\ups$ and
final state $\psis$, $P_1$, and $P_2$ denote the momenta of $\etac$ and $\psis$,  respectively.

As mentioned in Ref.~\cite{Sang:2011fw}, the exclusive decay of a bottomonium into double charmonia
involves the annihilation of a $b\bar{b}$ pair followed by the creation of two pairs of $c\bar{c}$.
One may guess that the generalization of the NRQCD factorization for the electromagnetic decay
or light-hadronic decay into this exclusive mode is possible.
So according to the NRQCD factorization formalism, the Lorentz scalar $A$ can be factorized as
\beq{\label{eq-nrqcd-factorization}}
A=\sqrt{2m_{\etac}2m_{\psis}}\sqrt{\langle {\cal O}\rangle_{\etac}\langle {\cal O}\rangle_{\psis}\langle {\cal O}\rangle_{\ups}}\,\,
c_0\PA{1+c_{2,1}\v2{\ups}+c_{2,2}\v2{\etac}+c_{2,3}\v2{\psis}+{\cal O}(v^4)},
\eeq
where $c_0$ and $c_2$ correspond to the SDCs, which can be calculated perturbatively
in power of $\as$.
For simplicity, we define shortcuts for the matrix elements
\begin{subequations}
\label{eq-matrixlement-1}
\begin{eqnarray}
\langle {\cal O}\rangle_{\ups}&=&|\langle0|\chi^\dagger\bm{\sigma}\cdot\bm{\eps}_1^*\psi|\ups(\eps_1)\rangle|^2,\\
\langle {\cal O}\rangle_{\etac}&=&|\langle\etac|\psi^\dagger\chi|0\rangle|^2,\\
\langle {\cal O}\rangle_{\psis}&=&|\langle\psis(\eps_2)|\psi^\dagger\bm{\sigma}\cdot\bm{\eps}_2\chi|0\rangle|^2,
\end{eqnarray}
\end{subequations}
and
\begin{subequations}
\label{eq-matrixlement-2}
\begin{eqnarray}
\v2{\ups}&=&\frac{\langle0|\chi^\dagger\bm{\sigma}\cdot\bm{\eps}_1^*\big(-\tfrac{i}{2}
\overleftrightarrow{\bf{D}}\big)^2\psi|\ups(\eps_1)\rangle}{m_b^2\langle0|\chi^\dagger\bm{\sigma}
\cdot\bm{\eps}_1^*\psi|\ups(\eps_1)\rangle},\\
\v2{\etac}&=&\frac{\langle\etac|\psi^\dagger\big(-\tfrac{i}{2}
\overleftrightarrow{\bf{D}}\big)^2\chi|0\rangle}{m_c^2\langle\etac|\psi^\dagger\chi|0\rangle},
\\
\v2{\psis}&=&\frac{\langle\psis(\eps_2)|\psi^\dagger\bm{\sigma}\cdot\bm{\eps}_2\big(-\tfrac{i}{2}
\overleftrightarrow{\bf{D}}\big)^2\chi|0\rangle}{m_c^2\langle\psis(\eps_2)|\psi^\dagger\bm{\sigma}\cdot\bm{\eps}_2\chi|0\rangle}.
\end{eqnarray}
\end{subequations}
In Eq.~(\ref{eq-nrqcd-factorization}), the bottomonium is normalized nonrelativistically on both sides;
in contrast, the charmonia are normalized relativistically on the left-hand side, but
 nonrelativistically on the right-hand side. The explicit factor $\sqrt{2m_{\etac}2m_{\psis}}$
is responsible for the discrepancy.

The SDCs are free of nonperturbative effects and can be obtained through the matching technique. To calculate
the SDCs, we are allowed to substitute the quarkonium states to free heavy quark pairs of the same quantum numbers as the corresponding hadrons.
Now $A$ and the matrix elements in  (\ref{eq-nrqcd-factorization}) can be calculated perturbatively; therefore,
the SDCs are readily derived. The spin-singlet and -triplet states of a free heavy quark pair can be extracted by employing the spin projectors in Ref.~\cite{Bodwin:2002hg}:
\begin{subequations}
\label{spin-projector}
\begin{eqnarray}
\Pi_3(p_0,\bar{p}_0)&=&\frac{1}{8\sqrt{2N_c}E(q_0)^2[E(q_0)+m_b]}
(/\!\!\!{p}_0+m_b)[\,/\!\!\!P_0\!+\!2E(q_0)]\,/\!\!\!\epsilon_1 (/\!\!\!\bar{p}_0-m_b),\\
\overline{\Pi}_1(p_1,\bar{p}_1)&=&\frac{1}{4\sqrt{2N_c}E(q_1)[E(q_1)+m_c]}
(/\!\!\!\bar{p}_1-m_c)\,\gamma_5[\,/\!\!\!P_1\!+\!2E(q_1)] (/\!\!\!{p}_1+m_c),\\
\overline{\Pi}_3(p_2,\bar{p}_2)&=&-\frac{1}{4\sqrt{2N_c}E(q_2)[E(q_2)+m_c]}
(/\!\!\!\bar{p}_2-m_c)\,/\!\!\!\epsilon_2^*[\,/\!\!\!P_2\!+\!2E(q_2)] (/\!\!\!{p}_2+m_c),
\end{eqnarray}
\end{subequations}
where a color-singlet factor $\tfrac{1}{\sqrt{N_c}}$ is included,
and $E(q_i)=\sqrt{m^2+{\bm q}_i^2}$ with $m=m_b$ for $i=0$ and $m=m_c$ for $i=1,2$.
In (\ref{spin-projector}), we take $P_i$ and $q_i$
to be the total momenta and half of the relative momenta of the quark pairs in the quarkonia,
where $i=0$ denotes
$\ups$, $i=1$ denotes $\etac$, and $i=2$ denotes $\psis$. Therefore, the momenta of the
corresponding quarks $p_i$ and antiquarks $\bar{p}_i$ in the hadrons are expressed as
\begin{subequations}
\label{kinematic-1}
\begin{eqnarray}
p_i&=&\frac{1}{2}P_i+q_i,\\
{\bar p}_i&=&\frac{1}{2}P_i-q_i.
\end{eqnarray}
\end{subequations}
$P_i$ and $q_i$ are chosen to be orthogonal: $P_i\cdot q_i=0$.

To obtain the next-to-leading-order (NLO) relativistic corrections for $S$-wave quarkonia,
we can expand the amplitude in powers of $q_i$, keep the quadratic terms
and make the following substitution
\begin{equation}
\label{eq-nlo1}
q_i^\mu q_i^\nu \to \bigg(-g^{\mu\nu}+\frac{P_i^\mu P_i^\nu}{4E(q_i)}\bigg)\frac{{\bm q}_i^2}{d-1},
\end{equation}
where $d$ is the dimension of spacetime.
In addition, the contributions from $E(q_i)$ can be readily derived by expanding
\begin{equation}
\label{eq-nlo2}
E(q_i)=\sqrt{m^2+{\bm q}_i^2}=m+\frac{{\bm q}_i^2}{2m}+{\cal O}(\frac{{\bm q}_i^4}{m^4}),
\end{equation}
where $m=m_b$ for $i=0$ and $m=m_c$ for $i=1,2$.

\section{the relativistic corrections to the SDCs\label{sec-amplitudedetail}}

In this section, we calculate the SDCs of $\ups\to \etac\psis$ accurate up to NLO relativistic
corrections following the techniques described in last section.
There are three channels which contribute, $\ups\to g^*g^*g^*\to \etac\psis$,
 $\ups\to g^*g^*\gamma^*\to \etac\psis$, and $\ups\to \gamma^*\to \etac\psis$.
In the subsequent three subsections, we will calculate the SDCs of the three channels separately.

\begin{figure}[ht]
\begin{center}
\includegraphics*[scale=0.7]{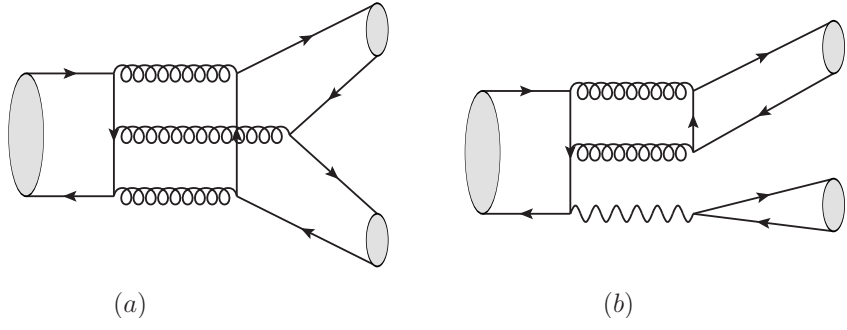}
\caption{The typical Feynman diagrams for $\Upsilon\to \eta_c J/\psi$, where Fig.~1(a) and Fig.~1(b) correspond to the channels $\Upsilon\to g^*g^*g^*\to \eta_c J/\psi$ and $\Upsilon\to g^*g^*\gamma^*\to \eta_c J/\psi$, respectively.}
\label{fig-1}
\end{center}
\end{figure}
\begin{figure}[ht]
\begin{center}
\includegraphics*[scale=0.8]{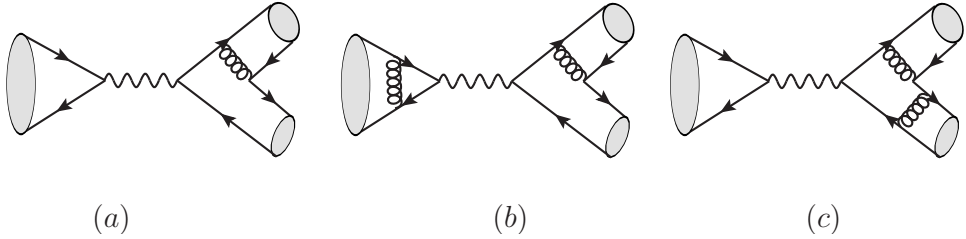}
\caption{The typical Feynman diagrams for $\Upsilon\to \gamma^*\to\eta_c J/\psi$. Fig.~2(a) corresponds to the tree-level diagram, and Fig.~2(b) and Fig.~2(c) correspond to the diagrams of the NLO radiative corrections from the initial state and final states, respectively.}
\label{fig-2}
\end{center}
\end{figure}
\subsection{$\ups\to g^*g^*g^*\to \etac\psis$}

The typical Feynman diagram is illustrated in Fig.~\ref{fig-1}(a), which is drawn by
using {\sc JaxoDraw}~\cite{Binosi:2008ig}. According to the Feynman rules,
we write down the amplitude of $b\bar{b}(p_0,\bar{p}_0)\to  g^*g^*g^*\to c\bar{c}(p_1,\bar{p}_1)+c\bar{c}(p_2,\bar{p}_2)$. The corresponding spin states for
the heavy quark pairs are readily obtained using
(\ref{spin-projector}), and the relativistic corrections are calculated by employing (\ref{eq-nlo1}) and (\ref{eq-nlo2}).
In our calculation, we use the {\sc Mathematica} package {\sc FeynArts}~\cite{Hahn:2000kx} to generate the Feynman diagrams and Feynman amplitude, {\sc FeynCalc}~\cite{Mertig:1990an} as well as {\sc FeynCalcFormLink}~\cite{Feng:2012tk} to implement the algebras of Dirac trace and Lorentz
indices contraction, and {\sc Fire}~\cite{Smirnov:2008iw,Smirnov:2014hma} and self-written programs~\cite{Feng:2012iq} to
make the tensor reduction. After some tedious work, we finally get the amplitude, accordingly,
the SDCs are obtained by removing the tensor and vectors in (\ref{eq-formfactor})
and producing the matching factor
\begin{equation}
\label{eq-operator-factor}
\sqrt{\frac{1}{2N_c}\frac{1}{2N_c4E(q_1)^2}\frac{1}{2N_c4E(q_2)^2}}.
\end{equation}
Since there are $E(q_i)$ in (\ref{eq-operator-factor}), we should further expand them by using (\ref{eq-nlo2}).

The SDCs are finally obtained analytically. As we will see in the next section, the channel $\ups\to g^*g^*g^*\to \etac\psis$ dominates $\ups$ decay into $\etac+\psis$; hence, it is both necessary and valuable for us to present the explicit expressions for these SDCs in the Appendix. Our LO SDCs agree with those in \cite{Jia:2007hy}.
The numerical predictions will be given in the following section.

\subsection{$\ups\to g^*g^*\gamma^*\to \etac\psis$}

The Feynman diagram is illustrated in Fig.~\ref{fig-1}(b).
For this channel, the $\psis$ is produced through the fragmentation of a photon.
At leading order in $\as$, the channel can be splitted into $\ups\to \etac\gamma^*$ and
$\gamma^*\to \psis$.  As discussed in Ref.~\cite{Bodwin:2007ga},
to reduce the theoretical uncertainty, the authors apply the VMD method~\cite{Bodwin:2006yd} to calculate
the fragmentation of the $\gamma^*$ to $\psis$ in studying the process
$e^+e^-\to \gamma^*\to \etac\gamma^*\to \etac\psis$. Similarly, here
we adopt the same treatment as that in Ref.~\cite{Bodwin:2007ga}.
To this end, we first calculate the amplitude of the subprocess
$\ups\to \etac\gamma^*$, then multiply the amplitude with a factor
\beq
\frac{g_{\psis}\sqrt{4\pi\alpha}}{m^2_{\psis}},
\eeq
where the effective coupling $g_{\psis}$ reads~\cite{Bodwin:2006yd}
\beq
g_{\psis}=\bigg(\frac{3m_{\psis}^3}{4\pi\alpha^2}\Gamma[\psis\to e^+e^-]\bigg)^{1/2},
\eeq
where $\Gamma[\psis\to e^+e^-]$ signifies the decay rate of the process $\psis\to e^+e^-$.

To get the SDCs, we apply the same techniques as in the last subsection. Finally, we obtain the analytic
expressions for the amplitude, and immediately also for the SDCs. Unfortunately, the expressions of these SDCs are much cumbersome,
therefore we do not attempt to  present them in this paper, and merely yield the numerical predictions in the next section.

\subsection{$\ups\to \gamma^*\to \etac\psis$}

The Feynman diagrams are illustrated in Fig.~\ref{fig-2}. To the accuracy considered in the current work,
the amplitude of this channel can be separated into two subprocesses, $\ups\to \gamma^*$ and $\gamma^*\to \etac\psis$. The NLO relativistic and radiative corrections to both subprocesses have been investigated.

For $\ups\to \gamma^*$, the amplitude accurate up to ${\cal O}(\as v^2)$ reads~\cite{Bodwin:2008vp}~\footnote{It can also be read from Ref.~\cite{Lee:2010ts} by taking $m_c= m_b$.}
\beq\label{eq-ups-gamma}
A(\ups\to \gamma^*)=A^{\rm Tree}(\ups\to\gamma^*)\bigg[1-\frac{2\as C_F}{\pi}
+\bigg(-\frac{1}{6}+\frac{\as C_F}{4\pi}(\frac{8}{9}+\frac{8}{3}\ln\frac{\mu_f^2}{m_b^2})\bigg)\langle v^2\rangle_{\ups}\bigg],
\eeq
where $\mu_f$ represents the factorization scale.
For $\gamma^*\to \etac\psis$, the amplitude with the photon energy the same as
that at B factories has been evaluated accurate up to ${\cal O}(\as v^2)$ corrections in~\cite{Dong:2012xx,Xi-Huai:2014iaa}.
In our process, since the virtual photon is produced through $\ups$ annihilation,
it is reasonable for us to take the invariant mass of $\gamma^*$ to be the mass of $\ups$.
After some effort, we are able to obtain the expression of the amplitude for the channel $\gamma^*\to \etac\psis$ accurate up to ${\cal O}(\as v^2)$ corrections, and immediately yield the amplitude for $\ups\to \gamma^*\to \etac\psis$ with the help of (\ref{eq-ups-gamma}). Our result for $\gamma^*\to \etac\psis$ with
the same energy as B factories agrees with Refs.~\cite{Dong:2012xx,Xi-Huai:2014iaa}, where
the subprocess $\gamma^*\to \etac\psis$ has been deeply investigated in ~\cite{Dong:2012xx,Xi-Huai:2014iaa}. For details, we refer the authors to Refs.~\cite{Dong:2012xx,Xi-Huai:2014iaa}.
In the following section, we will present the numerical predictions.

\section{numerical predictions and discussions\label{sec-prediction}}
In this section, we select input parameters,
and utilize the obtained analytic expressions
to make numerical predictions for the SDCs as well as the decay rates.

\subsection{Choosing the input parameters}

To get the decay rates, we should calculate the squared amplitude
and the phase space factor.
The squared amplitude is directly derived by multiplying the amplitude in (\ref{eq-formfactor}) with its complex conjugate. Sum over the polarizations for the final state $\psis$ and average the spin states for the initial state $\ups$, we get
\beq
\label{eq-A2}
\frac{1}{3}\sum_{spin}|{\cal A}|^2=\frac{2}{3}\PB{(P_1\cdot P_2)^2-P_1^2P_2^2}\times |A|^2,
\eeq
where, to maintain the gauge invariance, one must take $P_1^2=4E^2(q_1)$, $P_2^2=4E^2(q_2)$ and $P_1\cdot P_2=2\PB{E^2(q_0)-E^2(q_1)-E^2(q_2)}$.
On the other hand, the phase space factor does not affect the gauge invariance; therefore,
we use the physical masses of the quarkonia there~\cite{Bodwin:2008vp}
\beq\label{eq-phasespace}
\frac{1}{8\pi m_\ups^2}\lambda(m_\ups^2,m_{\psis}^2,m_{\etac}^2)^{1/2},
\eeq
where $\lambda$ is defined via $\lambda(a,b,c)\equiv a^2+b^2+c^2-2ab-2ac-2bc$.

We now specify our choices of the parameters in our computation.
We take the heavy quark pole masses as $m_c=1.4$ GeV and $m_b=4.6$ GeV. The masses of
quarkonia are taken from Ref.~\cite{Agashe:2014kda}:
\begin{subequations}
\label{parameter-mass}
\begin{eqnarray}
&&m_{\ups(1S)}=9.46030\ {\rm GeV},
m_{\ups(2S)}=10.02326\ {\rm GeV},
m_{\ups(3S)}=10.3552\ {\rm GeV},\\
&&m_{\etac(1S)}=2.9836\ {\rm GeV},
m_{\etac(2S)}=3.6394\ {\rm GeV}, m_{J/\psi}=3.096916\ {\rm GeV}.
\end{eqnarray}
\end{subequations}

The matrix elements for $J/\psi$ are taken from Ref.~\cite{Bodwin:2007fz},
\begin{eqnarray}
\label{eq-NRQCD-matrix-jpsi}
\langle {\cal O}\rangle_{J/\psi} =0.446\,{\rm
GeV}^3,\,\,\,\,\, \langle v^2\rangle_{J/\psi}=0.223,
\end{eqnarray}
which are fitted through the decay $J/\psi\to e^+e^-$ accurate through the relative order $v^2$,
and the matrix elements of $\etac$ are taken from Ref.~\cite{Guo:2011tz},
\begin{subequations}
\label{eq-NRQCD-matrix-etac}
\begin{eqnarray}
&&\langle {\cal O}\rangle_{\etac(1S)} =0.398\,{\rm
GeV}^3,\,\,\,\,\, \langle v^2\rangle_{\etac(1S)}=0.232,\\
&&\langle {\cal O}\rangle_{\etac(2S)} =0.202\,{\rm
GeV}^3,\,\,\,\,\, \langle v^2\rangle_{\etac(2S)}=0.255,
\end{eqnarray}
\end{subequations}
which are fitted through the decay $\etac(1S)\to 2\gamma$, $\etac(1S)\to {\rm LH}$,
and $\etac(2S)\to {\rm LH}$. In (\ref{eq-NRQCD-matrix-jpsi}) and (\ref{eq-NRQCD-matrix-etac}),
we have used the vacuum saturation approximation~\cite{Bodwin:1994jh} to
relate the production matrix elements with the decay ones.

In addition, we should also choose the values of the matrix elements for $\ups$,
which have been fitted in Refs.~\cite{Chung:2010vz,Chen:2011ph}.
Here, we cite the values in Ref.~\cite{Chung:2010vz},
\begin{subequations}
\label{eq-NRQCD-matrix-ups}
\begin{eqnarray}
&&\langle {\cal O}\rangle_{\ups(1S)} =3.069\,{\rm
GeV}^3,\,\,\,\,\, \langle v^2\rangle_{\ups(1S)}=-0.009,\\
&&\langle {\cal O}\rangle_{\ups(2S)} =1.623\,{\rm
GeV}^3,\,\,\,\,\, \langle v^2\rangle_{\ups(2S)}=0.09,\\
&&\langle {\cal O}\rangle_{\ups(3S)} =1.279\,{\rm
GeV}^3,\,\,\,\,\, \langle v^2\rangle_{\ups(3S)}=0.155,
\end{eqnarray}
\end{subequations}
which are fitted through the decay $\ups\to e^+e^-$.

In the channel $\ups\to g^*g^*\gamma^*\to \etac\psis$, we also need the decay rate for $J/\psi\to e^+e^-$,
 which is taken from the Ref.~\cite{Agashe:2014kda} as $\Gamma[J/\psi\to e^+e^-]=5.55$ keV.
Since the fine structure constant runs slightly with the energy scale,
we uniformly take $\alpha=\tfrac{1}{137}$. Finally, we choose the renormalization and factorization scales to be $\mu_R=\mu_f=m_b$, and
the strong coupling constant $\as(m_b)=0.22$~\cite{Jia:2007hy}.

\subsection{Numerical predictions for the SDCs}

In this subsection, we present the SDCs numerically by using the parameters selected in the previous subsection.
As mentioned, there are three channels which contribute to SDCs; hence,
we add superscripts in the SDCs to make the distinction, i.e., $c^{(i)}$, where $i=1,2,3$ corresponds
to the channels $\ups\to g^*g^*g^*\to \etac\psis$,
 $\ups\to g^*g^*\gamma^*\to \etac\psis$, and $\ups\to \gamma^*\to \etac\psis$, respectively.
 Substituting the parameters into the analytic expressions of the SDCs, we obtain
 for $\ups(1S)\to g^*g^*g^*\to \etac(1S)\psis$,
\begin{subequations}
\label{eq-numerical-result-1}
\begin{eqnarray}
c_0^{(1)}&=&(-2.16\times 10^{-5}-8.90\times
   10^{-5} i) \as^3\ {\rm GeV}^{-7}, \\
    c_{2,1}^{(1)}&=&(-5.64-1.83 i),\, c_{2,2}^{(1)}=(-1.14-0.90 i),\, c_{2,3}^{(1)}=(-0.99-0.27 i);
\end{eqnarray}
\end{subequations}
for $\ups(1S)\to g^*g^*\gamma^*\to \etac(1S)\psis$,
\begin{subequations}
\label{eq-numerical-result-2}
\begin{eqnarray}
c_0^{(2)}&=&(1.90\times 10^{-4}+7.79\times
   10^{-5} i)\alpha\as^2 \ {\rm GeV}^{-7},\\
 c_{2,1}^{(2)}&=&(-3.34-0.62 i) ,\, c_{2,2}^{(2)}=(-0.73+0.31 i),\,c_{2,3}^{(2)}=(0.19+0.02 i);
\end{eqnarray}
\end{subequations}
 and for $\ups(1S)\to \gamma^*\to \etac(1S)\psis$,
\begin{subequations}
\label{eq-numerical-result-3}
\begin{eqnarray}
c_0^{(3)}&=&(4.14\times 10^{-5}+6.78\times
   10^{-5}\as-7.08\times 10^{-5}\as i)\alpha\as\ {\rm GeV}^{-7},\\
 c_{2,1}^{(3)}&=&(-0.17-0.05\as),\,
 c_{2,2}^{(3)}=(0.52-1.60\as+2.50\as i),\\
  c_{2,3}^{(3)}&=&(0.35-0.89\as+1.76\as i).
\end{eqnarray}
\end{subequations}

To see the dependence on $\alpha$ and $\as$, we retain these two constants in the
SDCs explicitly.
From (\ref{eq-numerical-result-1}--\ref{eq-numerical-result-3}), we notice some characteristics. First,
$\ups\to g^*g^*g^*\to \etac\psis$ is the dominant channel for the SDCs and therefore also for the
decay rates. Second, the SDCs of the three channels are of quite different phase, in addition,
the phases of various relativistic corrections are also distinguishing.
The last is not the least, the relativistic corrections from $\ups$ are extraordinarily large,
so one may expect that this corrections will also contribute a lot to the decay rates
even though the corresponding matrix element is rather small.

\subsection{Numerical predictions for the decay rates}

Combining the squared amplitude (\ref{eq-A2}) with the phase space factor (\ref{eq-phasespace}),
we are able to evaluate the numerical predictions for the decay rates of various processes.
All the decay rates are listed in Table.~\ref{table-decayrate}.
\begin{table}
\caption{\label{table-decayrate}
The LO and NLO relativistic corrections to decay rates of $\ups\to \etac\psis$.
$\Gamma_{\rm LO}$ represents the decay rate at leading order in velocity expansion;
$\Gamma_{\rm NLO}^{(i)}$ with $i=0,1,2$ represent the relativistic corrections from $\ups$, $\etac$, and $\psis$,
respectively; $\Gamma_{\rm NLO}$ represents the total relativistic corrections.}
\begin{tabular}{c|c|c|c|c|c}
\hline
Processes & $\Gamma_{\rm LO}$ (eV) & $\Gamma_{\rm NLO}^{(0)}$ (eV) & $\Gamma_{\rm NLO}^{(1)}$ (eV) & $\Gamma_{\rm NLO}^{(2)}$ (eV) & $\frac{\Gamma_{NLO}}{\Gamma_{\rm LO}}$
\\
\hline
$\ups(1S)\to \etac(1S) J/\psi$& $0.41$ &0.03 &-0.28&-0.22 &$-115\%$
\\
\hline
$\ups(1S)\to \etac(2S) J/\psi$&$0.23$ & 0.02 &-0.17&-0.13 & $-122\%$
\\
\hline
$\ups(2S)\to \etac(1S) J/\psi$&$0.22$ & -0.18 &-0.15&-0.12 &$-204\%$
\\
\hline
$\ups(2S)\to \etac(2S) J/\psi$&$0.13$ & -0.11 &-0.10&-0.07 &$-211\%$
\\
\hline
$\ups(3S)\to \etac(1S) J/\psi$&$0.18$ & -0.25 &-0.12&-0.10 &$-263\%$
\\
\hline
$\ups(3S)\to \etac(2S) J/\psi$&$0.10$ & -0.15 &-0.08&-0.06 &$-270\%$
\\
\hline
\end{tabular}
\end{table}

In the table, we give the LO decay rates $\Gamma_{\rm LO}$ as well as the contributions
from the relativistic corrections. To see it clearly, we separately list the relativistic
corrections from different sources; i.e., $\Gamma_{\rm NLO}^{(i)}$ with
$i=0,1,2$ originates from $\ups$, $\etac$, and $\psis$, respectively.
From the table, we find that almost all the relativistic corrections are both significant
and negative, except for $\Gamma_{\rm NLO}^{(0)}$ in $\ups(1S)$ decay due to a
tiny and negative matrix element given in (\ref{eq-NRQCD-matrix-ups}).
Both the relativistic corrections from $\etac$ and $\psis$ are negative, which is quite different
with that in $\etac$ associated with $\psis$ production at B factories~\cite{He:2007te}.
The discrepancy may be accounted for by distinct Feynman diagram structure.

Another distinctive feature in the table is that the relativistic corrections from $\ups(2S,3S)$
are extraordinarily large, which even outnumber the corrections from charmonia, though the NLO NRQCD
matrix elements are smaller than these for charmonia. We may learn a lesson from this kind
of process and the example $\ups\to c\bar{c}gg$ in Refs.~\cite{Chen:2011ph,Chen:2012zzg}.
Despite a
rather small NRQCD matrix element $\langle v^2\rangle_{\ups}$,
it is not always reasonable to discard the relativistic corrections from $\ups$, which
actually have not been considered carefully in most references. Moreover, it is also important and
valuable to further determine the corresponding NRQCD matrix elements for $\ups$ through
independent approaches.

As mentioned in the introduction, $\ups\to \etac J/\psi$ is of a similar Feynman diagram structure
as $\psi\to \rho\pi$. We may expect there are also a relative large relativistic corrections for $\psi$
in $\psi\to \rho\pi$. Nevertheless, due to both large (even huge) and negative corrections,
the ``$\rho-\pi$'' puzzle is unlikely to be explained purely by the relativistic corrections, or
at least by the NLO corrections.

By combining all the corrections, we show the ratios of the decay rates of
the relativistic corrections to that at leading order in the Table.~\ref{table-decayrate}.
Our results indicate that the NLO relativistic corrections exceed the LO contributions for all the processes,
and what is worse, the corrections are negative, which renders the total decay rates
negative~\footnote{Since decay rates are expanded in powers of $v$ and $\as$,
and truncated at relative ${\cal O}(v^2)$, we obtain negative values.}
and therefore unpredictable at next-to-leading order. A resummation for the large relativistic corrections may be needed~\cite{Bodwin:2008vp,Lee:2010ts}, but is, however, out of the scope of this work.

In summary, we study the NLO relativistic corrections to $\ups$ exclusive decay into double
$S$-wave charmonia. We calculate the contributions from three channels
$\ups\to g^*g^*g^*\to \etac+\psis$, $\ups\to g^*g^*\gamma^*\to \etac+\psis$,
and $\ups\to \gamma^*\to \etac+\psis$.
Our results indicate, in spite of the small $\langle v^2\rangle_\ups$,
the relativistic corrections from $\ups$ are extraordinarily large, which suggests the
relativistic corrections from $\ups$ are not always negligible.
Moreover, we find that all the relativistic corrections to the decay rates are both
significant and negative, which indicates a poor convergence in
velocity expansion for this kind of process.

\begin{acknowledgments}
This work is supported by the National Natural Science
Foundation of China under
Grants No.~11447031, and No.~112755242, and by the Open Project Program of State Key Laboratory of Theoretical Physics, Institute of Theoretical Physics, under Grant No.~Y4KF081CJ1.
The work of W.-L. S. is, in part, supported by
the Natural Science Foundation of ChongQing under Grant No.~cstc2014jcyjA00029 and
by the Fundamental Research Funds for the Central Universities
under Grant No. SWU114003. The work of F.~F. is, in part, supported by the Fundamental Research Funds for the Central Universities.

\end{acknowledgments}
\appendix
\section{Analytic expressions of SDCs $c^{(1)}$ for the channel $\ups\to g^*g^*g^*\to\etac\jpsi$\label{appendix:tensor}}

For convenience, we rewrite the SDCs as
\begin{equation}
\label{appendix-sdc0}
c^{(1)}_{0}=\frac{\as^3}{6m_c^7}s_{0},\;\;\;c^{(1)}_{0}c^{(1)}_{2,1}=\frac{\as^3}{6m_c^7}s_{1},\;\;\;
c^{(1)}_{0}c^{(1)}_{2,2}=\frac{\as^3}{6m_c^7}s_{2},\;\;\;c^{(1)}_{0}c^{(1)}_{2,3}=\frac{\as^3}{6m_c^7}s_{3},
\end{equation}
and define

\begin{eqnarray}\label{appendix-def}
{\rm L1}&=&{\rm Re}\bigg[(-1+i \sqrt{3} \beta) \ln \frac{i (\beta-1)}{\sqrt{3}
   \beta-i}+(1+i \sqrt{3} \beta) \ln \frac{i (\beta+1)}{\sqrt{3}
   \beta+i}\bigg],\nn\\
   {\rm L2}&=&
   {\rm Re}\bigg[{\rm Li}_2\frac{(-i+\sqrt{3})
   \beta}{\sqrt{3} \beta+i}-{\rm Li}_2\frac{(-i+\sqrt{3})
   \beta}{\sqrt{3} \beta-i}\bigg],
   {\rm L3}=
   {\rm Li}_2\frac{1-\beta}{2},\nn\\
   {\rm L4}&=&{\rm Li}_2\frac{\beta-1}{\beta+1},{\rm L5}=
   {\rm Li}_2\frac{2 \beta}{\beta+1},{\rm L6}={\rm Li}_2\frac{\beta^2+\beta}{\beta-1},{\rm L7}=
   {\rm Li}_2\frac{\beta^2+\beta}{2-6
   r},\nn\\
   {\rm L8}&=&{\rm Li}_2\frac{\beta
   (\beta+1)^2}{4 (1-3 r)},{\rm L9}=
   {\rm Li}_2\frac{\beta-\beta^2}{\beta+1},{\rm L10}=
   {\rm Li}_2\frac{\beta-\beta^2}{6
   r-2},{\rm L11}={\rm Li}_2\frac{(\beta-1)^2 \beta}{12 r-4},\nn\\
   {\rm I1}&=&\ln2,{\rm I2}=\ln (1-\beta),{\rm I3}=\ln
   \frac{1-\beta}{1+\beta},{\rm I4}=\ln (1-3 r),   {\rm I5}=\ln
   \frac{4 (1-3 r)}{(\beta+1)^2},\nn\\
   {\rm A1}&=&\tan^{-1}(\sqrt{3} \beta),
\end{eqnarray}
where $r=\frac{m_c^2}{m_b^2}$, $\beta=\sqrt{1-4 r}$, and ${\rm Li_2}(x)$ represents the dilogarithm.

Utilizing the definitions in (\ref{appendix-def}), we present $s_0$, $s_1$, $s_2$, and $s_3$ as follows
\begin{eqnarray}\label{appendix-s0}
s_0&=&\frac{20 \pi
    (2 r-1) r^3 ({\rm I1}^2+{\rm I2}^2+2{\rm L3}+{\rm L4})}{27 \beta ^3}+\frac{5 \pi
   (10 r-3) r^3 {\rm I3}^2}{27 \beta ^3}\nn\\
   &&-\frac{5 \pi   (2 r-1) r [8
   r^2{\rm I3}+\beta  (2 r-1)]{\rm I2}}{27 \beta ^3}-\frac{5 \pi  (2
   r-1) r (\beta +8  r^2 {\rm I2}-8 r^2  {\rm I3}-2 \beta
   r) {\rm I1}}{27 \beta ^3}\nn\\
   &&+\frac{5 \pi   r \bigg[\beta
   -8 (2 {\rm I4}+i \pi +4) r^3+4 (\beta +i \pi +4) r^2-2 (2 \beta
   +3) r+1\bigg]{\rm I3}}{54 \beta ^3}\nn\\
   &&+\frac{40 \pi  r^4 }{27 \beta ^3}\bigg(\frac{2\pi}{3}{\rm A1}+ 2{\rm L2}+2{\rm L5}+{\rm L6}+{\rm L7}-{\rm L8}-{\rm L9}- {\rm L10}+{\rm L11}\bigg)\nn\\
    &&-\frac{5 \pi  r^2 \bigg[3 \beta ^3+2
   (2 \sqrt{3}-3 i) \pi  \beta  r+\pi ^2 r (2
   r-1)\bigg]}{81 \beta ^3},
\end{eqnarray}
\begin{eqnarray}\label{appendix-s1}
s_1&=&\frac{40  \pi  r^5 {\rm L1}}{27 (3 r-1) \beta ^5}+\frac{40
    \pi  r^4 ({\rm I4}-{\rm I5})}{27 \beta ^4}+\frac{10\pi (104 r^2-8 r-9)
   r^3  }{81 \beta ^5}(\frac{2\pi}{3}{\rm A1}+2 {\rm L2}+2 {\rm L5}+{\rm L6}+{\rm L7}\nn\\
   &&-{\rm L8}-{\rm L9}-{\rm L10}+{\rm L11})+\frac{10  \pi  (304
   r^2-206 r+37) r^3}{81 \beta ^5}({\rm I1}^2+{\rm I2}^2+2{\rm L3}+{\rm L4})\nn\\
   &&+\frac{5 \pi  (1120 r^2-634
   r+93) r^3 {\rm I3}^2}{162 \beta ^5}\nn\\
   &&-\frac{5 \pi} {972 (12 r^2-7
   r+1)^2 \beta }\bigg\{2 \pi ^2 (1-3
   r)^2 (304 r^2-206 r+37) r^3+4 \pi \bigg[8 (-405
   i+181 \sqrt{3}) r^3\nn\\
   &&+(3213 i-1532 \sqrt{3})
   r^2+6 (-177 i+88 \sqrt{3}) r-60 \sqrt{3}+117 i\bigg]
   \beta  r^3+(-14976 r^6+12888 r^5\nn\\
   &&-1162 r^4-2593 r^3+1286
   r^2-249 r+18) \beta\bigg\}+\frac{5
   \pi } {486\beta ^5 r}\bigg[24 (304 r^2-206 r+37)\nn\\
   && r^4({\rm I1}{\rm I3}-{\rm I1}{\rm I2}-{\rm I2}{\rm I3}) +(480
   r^5-44 r^4-200 r^3+235 r^2-84 r+9)({\rm I1}-{\rm I2}) \beta \bigg]\nn\\
   &&+\frac{5  \pi {\rm I3}} {486
   \beta ^5 r} \bigg\{\frac{\beta ^3} {(12 r^2-7 r+1)^2 (\beta
   +1)}
   \bigg[3456 (40+19 i \pi ) r^9-144 i (\pi  (114 \beta +727)\nn\\
   &&-48 i
   (7 \beta +33)) r^8+4 (14598 \beta +3 i \pi  (1839 \beta
   +5585)+34606) r^7+(-20974 \beta \nn\\
   &&-6 i \pi  (1873 \beta
   +3585)-25754) r^6+(-5731 \beta +24 i \pi  (107 \beta
   +144)-18253) r^5\nn\\
   &&+2 (4749 \beta -111 i \pi  (\beta +1)+8515)
   r^4-2 (2353 \beta +3431) r^3+(1225 \beta +1537) r^2\nn\\
   &&-3 (55 \beta
   +61) r+9 (\beta +1)\bigg]-12  r^4 (104 r^2-8 r-9){\rm I4}\bigg\},
\end{eqnarray}
\begin{eqnarray}\label{appendix-s2}
s_2&=&\frac{10 \pi  (13 r-1) r^3}{81 \beta ^5}\bigg(\frac{2\pi}{3}{\rm A1}+2 {\rm L2}+2{\rm L5}+{\rm L6}+{\rm L7}-{\rm L8}-{\rm L9}-{\rm L10}+{\rm L11}\bigg)\nn\\
&&+\frac{20 \pi
     (40 r^3-15 r+2) r^3({\rm I1}^2+{\rm I2}^2+2{\rm L3}+{\rm L4})}{81 \beta
   ^5} +\frac{5 \pi
    (120 r^3-32 r+5) r^3 {\rm I3}^2}{81 \beta
   ^5}\nn\\
   &&+\frac{5 \pi}{486
   \beta ^5}\bigg[ 48  (40 r^3-15
   r+2) r^3({\rm I1}{\rm I3}-{\rm I1}{\rm I2}-{\rm I2}{\rm I3}) +\beta
    (288 r^5-352 r^4-232 r^3\nn\\
    &&+224 r^2-48 r+3)({\rm I1}-{\rm I2})\bigg]-\frac{20 \pi
   r^4({\rm I4}-{\rm I5})}{27 \beta ^4}-\frac{20\pi r^5 {\rm L1}}{27 \beta^5 (3r-1)}\nn\\
   &&+\frac{5 \pi  {\rm I3}} {486 \beta ^5}\bigg\{-12  (13 r-1)
   r^3{\rm I4}-\frac{\beta ^3} {(\beta +1) (12 r^2-7
   r+1)^2}  \bigg[-3 (\beta +1)\nn\\
   &&-384 i (45 \pi -236 i)
   r^9+96 (290 \beta +15 i \pi  (3 \beta +11)+1402) r^8\nn\\
   &&+16 (-2523
   \beta -15 i \pi  (12 \beta -7)-5003) r^7+4 (3842 \beta -3 i \pi
   (95 \beta +527)+2110) r^6\nn\\
   &&+4 (1742 \beta +324 i \pi  (\beta
   +2)+5075) r^5+(-8805 \beta -12 i \pi  (27 \beta +35)-14809)
   r^4\nn\\
   &&+(3583 \beta +24 i \pi  (\beta +1)+4913) r^3-2 (367 \beta
   +439) r^2+(75 \beta +81) r\bigg]\bigg\}\nn\\
   &&-\frac{5
   \pi  r} {8748 \beta  (12 r^2-7
   r+1)^2}  \bigg[36 \pi ^2 (1-3 r)^2 (40 r^3-15 r+2)
   r^2\nn\\
   &&+4 \pi  \beta  (216 (70 \sqrt{3}-93 i) r^4-4
   (4036 \sqrt{3}-4401 i) r^3+(6468 \sqrt{3}-5931
   i) r^2\nn\\
   &&-18 (68 \sqrt{3}-59 i) r+100 \sqrt{3}-99
   i) r^2-9 \beta  (10560 r^6-9568 r^5+648 r^4+1983
   r^3\nn\\
   &&-824 r^2+123 r-6)\bigg],
\end{eqnarray}
and
\begin{eqnarray}\label{appendix-s3}
s_3&=&-\frac{100  \pi  r^5 {\rm L1}}{27 (3 r-1) \beta ^5}-\frac{100
    \pi  r^4({\rm I4}-{\rm I5})}{27 \beta ^4}+\frac{10  \pi (72 r^2+31
   r-1) r^3}{81 \beta ^5}\bigg(\frac{2\pi}{3}{\rm A1}+2 {\rm L2}+2{\rm L5}+{\rm L6}\nn\\
   &&+{\rm L7}-{\rm L8}-{\rm L9}-{\rm L10}+{\rm L11}\bigg)\nn\\
   &&+\frac{10  \pi
   (400 r^3-24 r^2-140 r+19) r^3({\rm I1}^2+{\rm I2}^2+2{\rm L3}+{\rm L4})}{81 \beta ^5}\nn\\
   &&+\frac{5
   \pi  (1200 r^3+72 r^2-358 r+55) r^3  {\rm I3}^2}{162 \beta
   ^5}-\frac{5 \pi}{8748 (12 r^2-7
   r+1)^2 \beta } \bigg[18 \pi ^2 (1-3 r)^2\nn\\
   && (400 r^3-24
   r^2-140 r+19) r^3+8 \pi  (540 (-93 i+70
   \sqrt{3}) r^4+(43524 i-40036 \sqrt{3}) r^3\nn\\
   &&+3
   (-4794 i+5291 \sqrt{3}) r^2-90 (-28 i+33
   \sqrt{3}) r+241 \sqrt{3}-234 i) \beta  r^3-9
   (52800 r^7\nn\\
   &&-57632 r^6+22872 r^5-7217 r^4+4017 r^3-1565
   r^2+279 r-18) \beta\bigg]\nn\\
   &&+\frac{5 \pi} {486 \beta
   ^5 r}\bigg[24
   (400 r^3-24 r^2-140 r+19) r^4 ({\rm I1}{\rm I3}-{\rm I2}{\rm I3}-{\rm I1}{\rm I2})+(1440 r^6-1904
   r^5\nn\\
   &&-4 r^4-100 r^3+307 r^2-99 r+9)({\rm I1}-{\rm I2}) \beta\bigg]\nn\\
   &&+\frac{5
   \pi {\rm I3} } {486 \beta ^5 r} \bigg\{\frac{\beta ^3}{(12
   r^2-7 r+1)^2 (\beta +1)} \bigg[1920 (236+45 i \pi )
   r^{10}\nn\\
   &&-96 i (3 \pi  (75 \beta +293)-2 i (725 \beta +3421))
   r^9+16 (12525 \beta +3 i \pi  (327 \beta -31)+25513) r^8\nn\\
   &&+8 i
   (537 \pi  \beta +12797 i \beta +3498 \pi +16349 i) r^7+2 (11339
   \beta -3 i \pi  (987 \beta +2003)+15363) r^6\nn\\
   &&+2 i (762 \pi  \beta
   +5240 i \beta +990 \pi +12631 i) r^5+(11041 \beta -114 i \pi
   (\beta +1)+20095) r^4\nn\\
   &&-4 (1411 \beta +2046) r^3+2 (716 \beta
   +887) r^2-18 (10 \beta +11) r+9 (\beta +1)\bigg]\nn\\
   &&-12  r^4 (72 r^2+31
   r-1){\rm I4}\bigg\}.
\end{eqnarray}



\end{document}